\documentclass[a4paper,11pt]{article}
\usepackage{pos}

\title{SU(5) aGUT: a minimal asymptotic grand unification model}
 \ShortTitle{Minimal SU(5) aGUT}

\author*[a,b]{Giacomo Cacciapaglia}
\author[c]{Ammar Abdalgabar}
\author[d]{Corentin Cot}
\author[e]{Alan S. Cornell}
\author[a,b]{Aldo Deandrea}
\author[f]{Mohammed Omer Khojali}

\affiliation[a]{Institut de Physique des 2 Infinis de Lyon (IP2I), UMR5822, CNRS/IN2P3,  F-69622 Villeurbanne Cedex, France}
\affiliation[b]{University of Lyon, Universit\'e Claude Bernard Lyon 1, F-69001 Lyon, France}
\affiliation[c]{University of Hafr Al Batin, college of Science, department of physics, Hafr Al Batin 39524, Kingdom of Saudi Arabia}
\affiliation[d]{Laboratoire de Physique des 2 Infinis (IJCLab), Université Paris-Saclay, Orsay, France}
\affiliation[e]{Department of Physics, University of Johannesburg, PO Box 524, Auckland Park 2006, South Africa}
\affiliation[f]{Department of Physics, University of Khartoum, PO Box 321, Khartoum 11115, Sudan}

\emailAdd{g.cacciapaglia@ipnl.in2p3.fr}

\abstract{We present a new grand unification paradigm, where gauge couplings do not need to be equal at any given scale, instead they run towards the same fixed point in the deep ultraviolet. We provide a concrete example based on SU(5) with a compactified extra space dimension. By construction, fermions are embedded in different SU(5) bulk fields, hence baryon number is conserved and proton decay is forbidden. The lightest Kaluza-Klein tier consists of stable states, providing an asymmetric Dark Matter candidate via their baryonic charges, with a mass of 2.4 TeV. The model features an interesting and predictive flavour structure.}

\FullConference{%
  41st International Conference on High Energy physics - ICHEP2022\\
  6-13 July, 2022\\
  Bologna, Italy
}

\begin{document}
\maketitle

\section{Introduction}

The Standard Model of particle physics (SM) consists of a quantum field theory constructed around the gauge principle, i.e. the presence of a local invariance under a symmetry transformation. Experimental evidence has shown that the gauge symmetry underlying the SM can be described in terms of a semi-simple Lie group: SU(3)$_c \times$ SU(2)$_L \times$ U(1)$_Y$. Each group factor corresponds to gauge interactions with distinct strength, hence characterised by different coupling constants $g_i$, $i=1,2,3$.
However, this structure is not static and unchangeable, as proven by two characteristics of the SM itself. On the one hand, coupling constants evolve with the energy of the physical processes, following the renormalisation group approach to quantum corrections \cite{Wilson:1973jj}. In the SM, with increasing energy the three gauge coupling values tend to approach each other around energy scales of $10^{12} \div 10^{18}$~GeV. On the other hand, the gauge structure itself is different at different scales. In the SM, below about $100$~GeV the gauge symmetry is reduced to SU(3)$_c \times$ U(1)$_{\rm em}$, following the spontaneous breaking of the electroweak scale. This phenomenon is described in the SM in terms of the Higgs sector \cite{Higgs:1964pj}, i.e. a scalar field that develops a non-zero vacuum expectation value (VEV). 

These features of the SM inspired the idea that, at high scales, the SU(3)$_c \times$ SU(2)$_L \times$ U(1)$_Y$ gauge structure may be replaced by a simple unified Lie group $\mathcal{G}_{\rm GUT}$, with a single gauge coupling. Two minimal possibilities exist following the fact that the fermion field in a SM family can be grouped in complete representations of SU(5) \cite{Georgi:1974sy}, or of SO(10) \cite{Fritzsch:1974nn} if right-handed neutrinos are included. This inevitably leads to the generation of proton decay operators at low energies, implying that grand unification theories (GUTs) can only occur at energy scales above $10^{16}$~GeV. Note that the existence of a single gauge coupling at high scales in traditional GUTs implies that the SM gauge couplings must undergo precision unification, i.e. be identical at a given energy scale, where the breaking of $\mathcal{G}_{\rm GUT}$ by a Higgs mechanism must occur. This feature has tremendous implications on the GUT model building. Firstly, new states are required at an intermediate scale to drive the running of the SM gauge couplings towards precision unification: this is usually achieved by introducing supersymmetry at the TeV scale. Secondly, fermions must live in complete representation of $\mathcal{G}_{\rm GUT}$, leading to inevitable proton decay. Thirdly, $\mathcal{G}_{\rm GUT}$ must be broken and appropriate Yukawa couplings generated, hence requiring the presence of a non-trivial set of scalar fields in large representations of the unified gauge group. This last point can be problematic, especially for supersymmetric GUTs: in fact, large representations impact the running of the unified gauge coupling, implying the presence of Landau poles right above the unification scale. Hence, the validity of traditional supersymmetric GUTs is very limited \cite{Bajc:2016efj}.

In this paper we put forward a novel approach to grand unification, where precision gauge coupling unification is not necessary. The SM gauge couplings, in fact, may run towards a fixed points at high scales, as occurring in Ultra-Violet safe theories \cite{Litim:2014uca}. Hence, instead of being equal at a finite energy scale, the gauge couplings run asymptotically towards the same UV fixed point, hence the name ``aGUT''. The ``a'' here stands for asymptotic, but also refers to the important fact that unification never really occurs. Namely, a unified gauge symmetry is never realised but emerges, approximately, at high scales. One key feature of aGUTs is that only one scale is needed, where some new dynamics is introduced to alter the running of the gauge couplings. This kind of unification can be achieved in theories with a large number of massive fermions, provided an intermediate Pati-Salam unification is achieved \cite{Molinaro:2018kjz}. In this work, we will follow the route of extra dimensional theories, pioneered in Ref.~\cite{Dienes:2002bg}. In theories with one compact extra space-like dimension, an effective 't Hooft coupling can be defined, including the effect of the Kaluza-Klein modes, namely $\tilde \alpha = \mu R\ \alpha$, where $\alpha = g^2/4\pi$ and $\mu$ is the renormalisation energy. In fact, $\mu R$ counts how many states lie below the renormalisation scale, where $R$ is the radius of the extra dimension that sets the scale of new physics in the model. Hence, the running of the 't Hooft coupling can be written as
\begin{equation} \label{eq:RGE}
    2 \pi \left( \tilde{\alpha} + \frac{d \tilde{\alpha}}{d \ln \mu} \right) = b_5\ \tilde{\alpha}^2\,, \qquad \tilde{\alpha}_{\rm UV} = -\frac{2\pi}{b_5}\,.
\end{equation}
In the above equation, the linear term in the left-hand side comes from the explicit energy dependence of the 't Hooft coupling, while $b_5$ is the one-loop coefficient from the 5D bulk fields. Thanks to the first term, an attractive fixed point for $\tilde \alpha \to \tilde{\alpha}_{\rm UV}$ emerges as long as $b_5 < 0$.
Note that this kind of 5D theories are well defined at all scales, leading to definition of renormalisable theories \cite{Gies:2003ic,Morris:2004mg} notwithstanding the fact that the 5D gauge coupling carries mass dimension.

In Ref.~\cite{Cacciapaglia:2020qky} we constructed the first explicit aGUT model in 5D, based on the minimal gauge group SU(5). One important model building constraint is due to the running of the Yukawa couplings, for which the UV fixed points are not attractive. Hence one needs to make sure that their running from the low energy SM values do not generate a Landau pole \cite{Abdalgabar:2017cjw}. This can rule out some possibilities, as for instance the minimal SO(10) aGUT \cite{Khojali:2022gcq}.
An interesting feature of these models is that Yukawa couplings are not required to unify, hence avoiding proton decay and allowing the extra dimensional scale $m_{\rm KK} \equiv 1/R$ to be as low as the TeV scale. 

\begin{figure*}[tb!]
\begin{centering}  \includegraphics[scale=0.5]{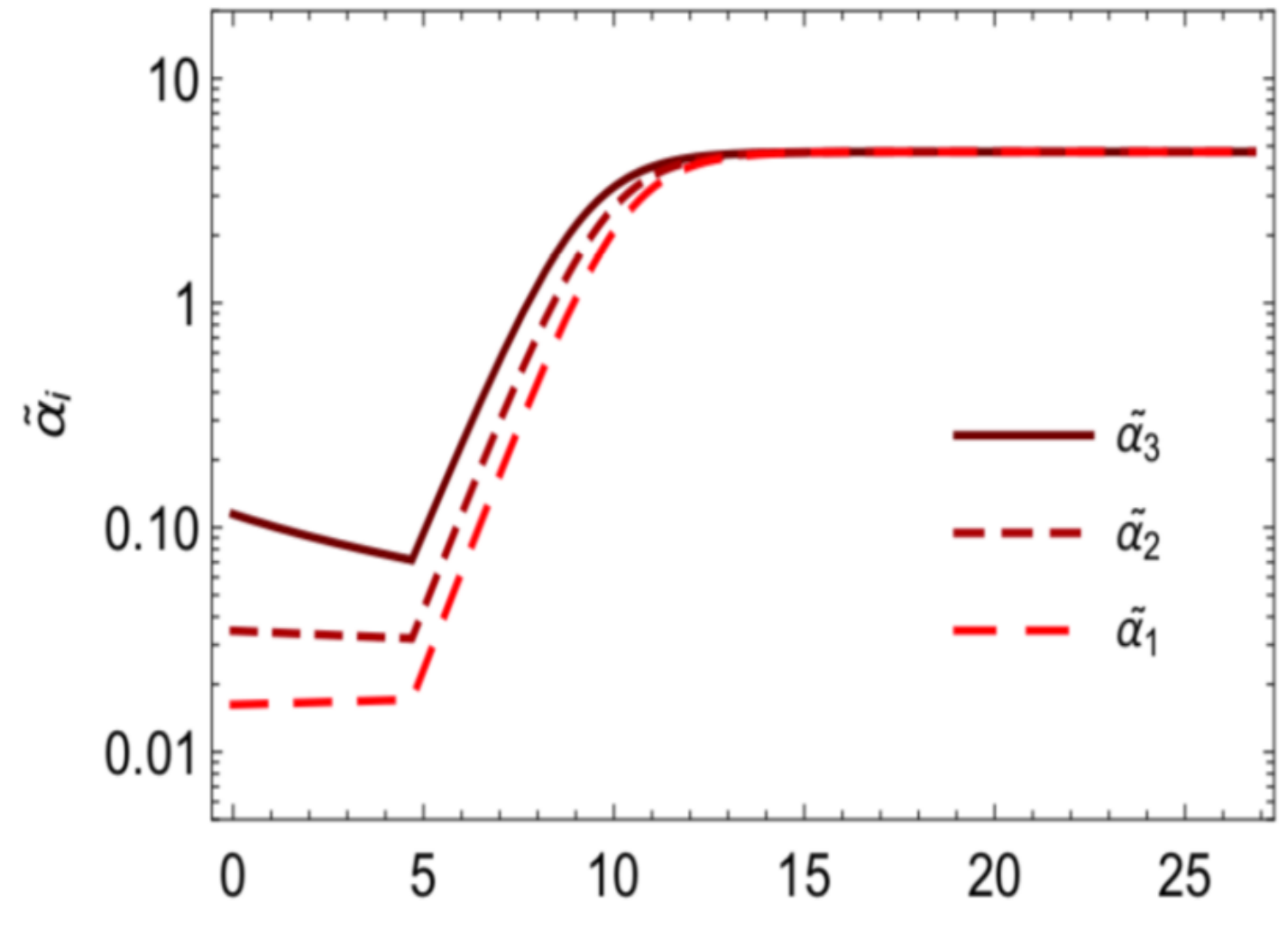}
\includegraphics[scale=0.55]{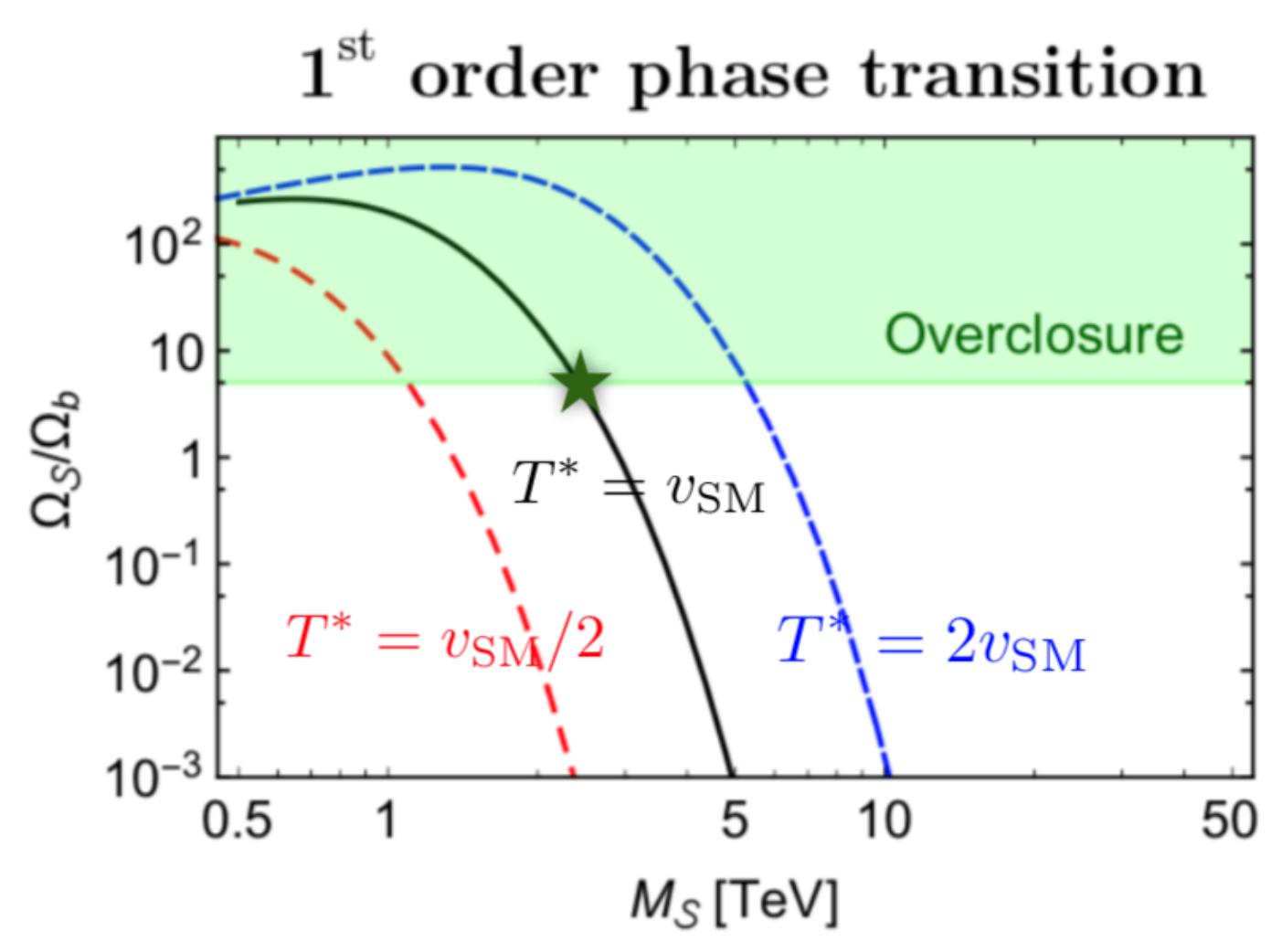} 
\end{centering}
    \caption{Left: schematic running of the 5D 't Hooft gauge couplings as a function of $t=\ln \mu/m_Z$. Right: estimate of the Indalo relic density from baryogenesis as a function of the Dark Matter mass. Plots adapted from Ref.~\cite{Cacciapaglia:2020qky}.}
    \label{fig}
\end{figure*}

\section{Minimal SU(5) aGUT and its flavour structure}

The minimal SU(5) aGUT model \cite{Cacciapaglia:2020qky} is based on a 5D background compactified on $S^1/\mathbb{Z}_2\times \mathbb{Z}_2'$. The two parities can be chosen in such a way that SU(5) is broken to the SM on one boundary, via $\mathbb{Z}_2$, while the second parity leaves the bulk group unbroken. A generic feature of 5D models is that, due to the SU(5) breaking, different components of an SU(5) multiplet have different parities. Henceforth, it is not possible to embed the SM fermion zero modes into the same multiplets, as it is usually done in traditional GUT model building. For instance, if we consider a bulk $\bar{5} \subset (b_R^c, l_L)$, choosing the parities that ensure a left-handed zero mode for $l_L$ implies that the $b_R^c$ component has no zero mode. The solution is to add a $5$ with flipped boundary conditions, such that a right-handed zero model is present for the $b_R$ component, and none for the leptons. This mechanism implies that: (a) the representation content of a bulk family is doubled with respect to traditional GUTs; (b) SM fermions are embedded in different bulk fields. In Table~\ref{tab:2} we list the minimal set of bulk fields with the appropriate parities assigned to the SM components. Note that the parities refer to the left-handed spinors, such that $(+,+)$ implies a left-handed zero mode, while $(-,-)$ implies a right-handed zero mode. We see that only the right handed up-type quarks $t_R$ and the right-handed charged leptons $\tau_R$ belong to the same bulk field. The SM Higgs, instead, is embedded in a bulk $5$, where the coloured component has no zero mode, hence solving the doublet-triplet splitting problem.

With this bulk field content, the 5D beta function reads
\begin{equation}
    b_5 = -\frac{52}{3} + \frac{16}{3} n_g\,,
\end{equation}
where $n_g$ is the number of bulk generations. We see that $b_5 < 0$, as required by the presence of fixed points in Eq.~\eqref{eq:RGE}, requires $n_g \leq 3$.
The running of the 't Hooft gauge couplings is shown in the left panel of Fig.~\ref{fig}, where the kink in the running indicate the mass scale $m_{\rm KK}$ where the extra dimensional effects kick in. The gauge couplings approach the same value above the compactification scale, but they are never identical.

\begin{table}[ht]
\begin{center}
\begin{tabular}{ |p{1.6cm}||p{1cm}||p{1.6cm}|p{1.2cm}|p{1cm}|p{1cm}|p{1cm}| }
 \hline
 Multiplets    & Fields  & SM gauge &  $(\mathbb{Z}_2, \mathbb{Z}_2')$ & Zero mode? & L & B\\
 \hline
 $\psi_{\overline{5}}$ & $B_R^c$ & $(\bar{3}, 1)_{1/3}$ & $(-,+)$ & - & 1/2 & 1/6\\
  &  \fbox{$l_L$} & $(1,2)_{-1/2}$ & $(+,+)$ & $\surd$ & 1 & 0\\
 \hline
 $\psi_{5}$ & \fbox{$b_R$} & $(3,1)_{-1/3}$ & $(-,-)$ & $\surd$ &  0 & 1/3 \\
 &  $L_L^c$ & $(1,2)_{1/2}$ & $(+,-)$ & - &  -1/2 & 1/2\\
 \hline
  $\psi_{10}$ & $T_R^c$ & $(\bar{3}, 1)_{-2/3}$ & $(-,+)$ & - & 1/2 & 1/6\\
  & $\mathcal{T}_R^c$ & $(1,1)_{1}$ & $(-,+)$ & - & -1/2 & 1/2\\
  & \fbox{$q_L$} & $(3,2)_{1/6}$ & $(+,+)$ & $\surd$ & 0 & 1/3\\
  \hline
  $\psi_{\overline{10}}$ & \fbox{$t_R$} &  $(3,1)_{2/3}$ & $(-,-)$ & $\surd$ & 0 & 1/3\\
  & \fbox{$\tau_R$} & $(1,1)_{-1}$ & $(-,-)$ & $\surd$ & 1 & 0\\
  & $Q_L^c$ & $(\bar{3}, 2)_{-1/6}$ & $(+,-)$ & - & 1/2 & 1/6\\
 \hline
 $\psi_1$ & $N$ & $(1,1)_0$ & $(-,-)$ & $\surd$ & $1$ & $0$ \\ 
\hline
 $\psi_1'$ & $\mathcal{S}$ & $(1,1)_0$ & $(+,-)$ & - & $-1/2$ & $1/2$ \\ 
 \hline
 $\phi_5$ & $H$ & $(3,1)_{-1/3}$ & $(-,+)$ & - & 1/2 & -1/6\\
 & \fbox{$\phi_h$} & $(1,2)_{1/2}$ & $(+,+)$ & $\surd$ & 0 & 0\\
 \hline
\end{tabular} \end{center}
\caption{List of 5D fermions and scalar fields, decomposed in SM components. We also list the SM quantum numbers, parities, lepton (L) and baryon (B) numbers associated to the components. Parities refer to left-handed spinors. The fields with zero modes, corresponding to the SM, are highlighted in a box.}\label{tab:2}
\end{table}

As the SM fermions are embedded in different bulk fields, the Yukawa couplings do not unify. This is due to the fact that one can write independent SU(5)-invariant couplings for the charged fermions, as follows: 
\begin{equation}
    \mathcal{L}_{\rm Yuk} = - \sqrt{2} Y_\tau\ \bar{\psi}_{\bar{5}}\ \psi_{\bar{10}}\ \phi_5 - \sqrt{2} Y_b\ \bar{\psi}_5\ \psi_{10}\ \phi_5^\ast - \frac{1}{2} Y_t\ \bar{\psi}_{\bar{10}}\ \psi_{10}\ \phi_5 + \mbox{h.c.} 
\end{equation}
Note that in Table~\ref{tab:2} we also included two singlets, to play the role of right-handed neutrinos and allow to generate neutrino masses. A consequence of the Yukawa couplings above is that baryon and lepton numbers emerge as conserved accidental symmetries, as in the SM. The charge assignment is reported in Table~\ref{tab:2}. Note that the components that do not have zero modes feature unusual charges, namely half of the usual baryon and lepton numbers of quarks and leptons. This implies that such states, whose lowest mode has mass $m_{\rm KK}$, cannot decay into SM states, hence leading to accidental stability. In Ref.~\cite{Cacciapaglia:2020qky}, these states were dubbed ``Indalo'', from a Zulu word that means creation or nature. The lightest such states, i.e. the singlet $\mathcal{S}$, can play the role of Dark Matter and an asymmetric relic density will be produced together with baryogenesis. In the right plot of Fig.~\ref{fig} we show the predicted relic density as a function of the mass, indicating that the asymmetric relic can saturate the Dark Matter one for $m_{\rm KK} \approx 2.4$~TeV. Larger masses are also allowed as long as a dominant thermal component is present.

The Yukawa couplings also have a distinct flavour structure. By diagonalising the Yukawa matrices, one can prove that two unitary matrices survive: one, $V_C$, corresponds to the CKM matrix in the SM, while another one, $V_I$, encodes new physics effects. In terms of SM and Indalo fermions, they appear as \cite{Khojali:wip}:
\begin{eqnarray}
    Y_b : && \bar{b}\cdot Y_b\cdot q\ \phi_h^\ast - \bar{L}^c\cdot Y_b\cdot q\ H^\ast - \bar{L}^c\cdot Y_b\cdot \mathcal{T}^c \phi_h^\ast + \bar{b}\cdot Y_b\cdot T^c\ H^\ast \,, \nonumber \\
    Y_\tau : &&  \bar{\tau}\cdot Y_\tau\cdot l\ \phi_h^\ast + \bar{Q}^c\cdot Y_\tau\cdot l\ H^\ast - \bar{Q}^c\cdot Y_\tau\cdot B^c\ \phi_h^\ast + \bar{t}\cdot Y_\tau\cdot V_I^T\ B^c\ H^\ast\,, \nonumber \\
    Y_t : && \bar{t}\cdot Y_t\cdot V_C\cdot q\ \phi_h + \bar{t}\cdot Y_t\cdot V_C\cdot \mathcal{T}^c\ H + \bar{\tau}\cdot V_I\cdot Y_t\cdot V_C\cdot T^c\ H + \bar{Q}^c \cdot V_I\cdot Y_t\cdot T^c\ \phi_h + \nonumber \\
    && + \bar{Q}^c\cdot V_I\cdot Y_t \cdot V_C\cdot q\ H\,. 
\end{eqnarray}
The new flavour matrix $V_I$ appears in the top sector, but also involves leptons via the right-handed zero modes. Hence, loops of the Indalo quarks and $H$ can generate lepton flavour violation. As an example, contribution to decays via a photon can be estimated as
\begin{equation}
    \frac{\mbox{BR}(l\to l'\gamma)}{\mbox{BR} (l\to l'\nu\nu)} = 8.4 \times 10^{-8}\ \left( \frac{2.4~\mbox{TeV}}{m_{\rm KK}} \right) \left| (V_I)^{lt} (V_I^\dagger)^{tl'} + \mathcal{O} (m_q^2/m_t^2) \right|^2\,,
\end{equation}
where $m_q$ is the mass of light generations. The strongest experimental bound comes from muon decays, $\mbox{BR} (\mu\to e \gamma) < 4.2\times 10^{-13}$, however is suffices to have mixing of the third generation to the lighter ones of the order of $10^{-3}$, like in the CKM, to safely suppress these effects. A more thorough analysis of flavour  in this model is under way \cite{Khojali:wip}.

\section{Conclusions and outlook}

We have presented the first concrete model in 5 dimensions implementing asymptotic grand unification, via the presence of attractive UV fixed points for the gauge coupling running. The model is based on a minimal SU(5) bulk gauge symmetry, broken via boundary conditions. The fermion embedding into different fields, required by the boundary conditions, implies conservation of baryon and lepton numbers. As a consequence, proton decay is avoided and the lightest KK mode is accidentally stable. This provides a natural asymmetric Dark Matter candidate. Furthermore, the flavour structure of the Yukawa couplings is well defined, leading to a new flavour mixing matrix in addition to the CKM one.

This example shows the many advantages of the aGUT approach to grand unification. Furthermore, the model building requirements are very restrictive, hence limiting the number of viable models much more than in standard GUTs.

\end{document}